\documentclass{eds2019}
\usepackage{amsmath}
\usepackage{amssymb}
\usepackage{xcolor}

\def\be{\begin{equation}}
\def\ee{\end{equation}}
\def\bea{\begin{eqnarray}}
\def\eea{\end{eqnarray}}

\newcommand{\Photo}{}

\begin{document}
\vspace*{4cm}
\title{Recent IceCube Measurements Using High Energy Neutrinos
}

\author{ Hans Niederhausen\footnote{email: \protect\url{hans.niederhausen@tum.de}} for the IceCube Collaboration\footnote{\protect\url{http://icecube.wisc.edu}} }

\address{Technische Universit\"at M\"unchen\\ Physik-Department, James-Franck-Str. 1, D-85748 Garching, Germany}

\maketitle\abstracts{
The IceCube Neutrino Observatory, located at the geographic South Pole, is a
Cherenkov detector that continuously monitors a cubic kilometer of 
instrumented glacial ice for neutrino interactions in the sub-TeV to 
EeV energy range. Its primary design goal is the study of powerful 
astrophysical objects that could act as natural particle accelerators and 
thus as sources of (ultra) high energy cosmic rays - in short: 
to do neutrino astronomy. IceCube 
has discovered a diffuse flux of high energy astrophysical neutrinos 
consistent with an extra-galactic origin. In addition the IceCube Collaboration recently 
obtained evidence for neutrino emission from the direction of the 
blazar TXS 0506+056, making it the first potentially identified source of 
high energy cosmic rays. IceCube also 
contributes to fundamental particle physics through the study of 
neutrino interactions at large energies. In this talk I
present recent results and measurements of high energy neutrinos with IceCube.}

\section{Introduction}

Many observed phenomena in the high energy, non-thermal universe are not well understood. One of the biggest unresolved questions is the origin and production mechanism of ultra-high energy cosmic rays. These protons and heavier nuclei have been observed with ground based observatories to have energies reaching beyond $10^{20}\,\textnormal{eV}$ \cite{Fenu2017,ivanov2019}. Extracting information about possible sources or source classes from these observations of charged cosmic rays is complicated, because these particles, depending on energy and atomic number, may experience significant magnetic deflection and rapid energy loss in the intergalactic medium. Secondary $\gamma$-rays and neutrinos, produced when cosmic-rays interact with photons or gas in or near cosmic accelerators, can point back to acceleration sites and thus might prove crucial in solving the cosmic ray puzzle. Neutrinos are of special interest because, unlike $\gamma$-rays, they cannot arise from purely leptonic processes. Thus, high-energy neutrinos provide definitive evidence of hadronic particle acceleration in cosmic ray sources \cite{Halzen2002}. Furthermore, neutrinos are the only astrophysical messenger particle for which the universe remains transparent over cosmological distances at the highest energies \cite{Greisen1966,Zatsepin1966,Gould1967,Nikishov1961}. Well-motivated neutrino source candidates include active galactic nuclei \cite{Dermer2014,Murase2017}, blazars in particular \cite{Padovani2015}, and gamma-ray bursts \cite{Waxman1997a,Tamborra2015}. Besides their obvious role in neutrino and multi-messenger astronomy, high energy cosmic neutrinos are also promising probes of fundamental particle interactions far beyond the reach of man-made particle accelerators \cite{Ackermann2019}. Because the expected flux of cosmic neutrinos and the neutrino-nuclueon interaction cross section are small, the measurements require very large volume detectors ($\gtrsim 1\,\mathrm{km}^{3}$) \cite{Halzen2010}.\\
The IceCube Neutrino Observatory is the largest neutrino telescope currently in operation. It is located at the geographic South Pole and consists of 86 cables that each hold 60 Digital Optical Modules (DOMs). The DOMs, deployed in the South Pole ice sheet at depths between $1450\,\mathrm{m}$ and $2560\,\mathrm{m}$, instrument one cubic kilometer of glacial ice. Each DOM contains a $10\mathrm{''}$ photomultiplier tube and supporting electronics, including 12 LEDs for in-situ calibration. Observed light signals that satisfy simple threshold and local coincidence requirements are digitized within each DOM and sent to the IceCube Laboratory on the surface for further processing \cite{Aartsen2017d}. IceCube detects the Cherenkov light emitted by charged secondary particles, produced in deep inelastic interactions between neutrinos and nucleons in the ice, or by atmospheric muons created when primary cosmic-rays collide with the nucleons in Earth's atmosphere. Muons leave a track-like signature in the detector which enables directional reconstruction to better than $1^\circ$ \cite{IceCubeCollaboration2019}. A cascade signature arises from neutrino induced particle showers via charged current interactions (CC) of $\nu_e$ and $\nu_\tau$, and neutral current interactions (NC) of all neutrino flavors. For cascades contained inside IceCube's fiducial volume, the energy deposit can be reconstructed with a resolution of $\sim 10\%$ and its arrival direction to $\sim 10^\circ$ \cite{IceCubeCollaboration:2014}.
Several sources of background events are present in IceCube data. The detector observes atmospheric muons at a rate of $\sim2.7\,\mathrm{kHz}$ from the Southern Sky \cite{Aartsen2017d} which are a significant background in the study of cosmic neutrinos. Another source of background events are neutrinos produced in the same air showers as these atmospheric muons (trigger rate $\sim\mathrm{mHz}$) \cite{Aartsen2017d}. At energies of interest for astrophysical searches ($\gtrsim \mathrm{TeV}$) conventional atmospheric neutrinos stem primarily from kaon decays ($K^{\pm},\,K^{0}_{L}$) and thus consist predominantely of $\nu_\mu$ and $\bar{\nu}_\mu$. Due to their relatively long lifetime kaons lose energy prior to decaying. Thus, the energy spectrum of conventional atmospheric neutrinos ($\sim E^{-3.7}$) is steeper than the primary cosmic-ray spectrum ($\sim E^{-2.7}$). The atmospheric neutrino flux is strongest for horizontal trajectories because of the increased atmospheric depth compared to vertical directions \cite{Gaisser2002}. At higher energies towards $\sim 100\,\mathrm{TeV}$ one expects heavy mesons involving charm quarks (e.g. D-mesons) to dominate the production of atmospheric neutrinos. Because these heavy meson decay rapidly, this prompt atmospheric neutrino flux has energy and declination dependence similar to the primary cosmic ray flux and thus appears isotropic. Furthermore, the prompt flux has equal contributions from electron and muon (anti)neutrino flavors \cite{Gaisser2002}.

\section{Identifying Astrophysical Neutrinos in IceCube}
\begin{figure}
  \centering
    \includegraphics[width=0.9\textwidth]{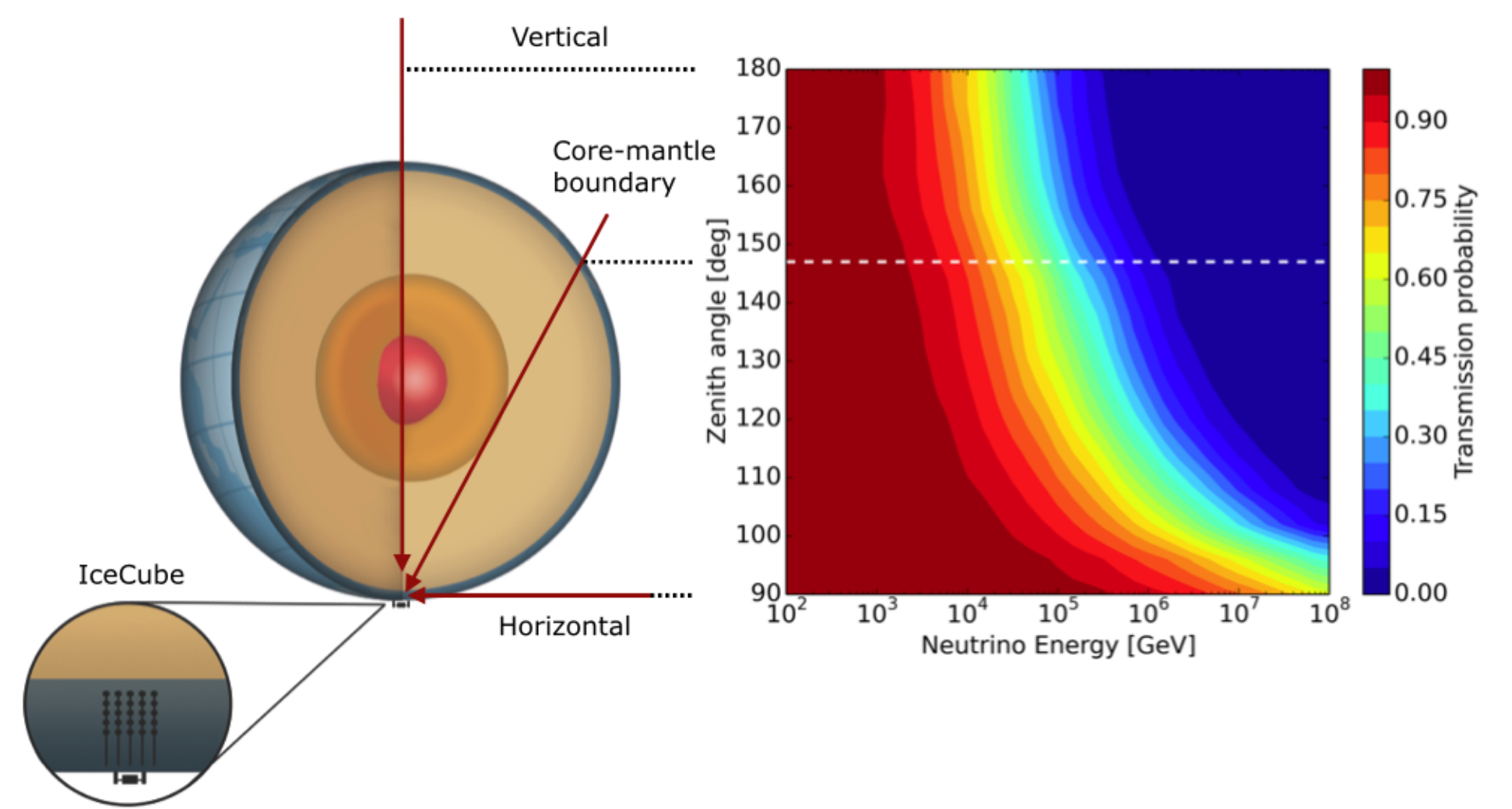}
      \caption{\textbf{Neutrino transmission through Earth:} transmission probability as function of zenith angle and energy. Absorption of neutrinos increases with the amount of material traversed and with neutrino energy.}
      \label{fig:absorption}
\end{figure}
A cosmic origin of neutrino events in IceCube can be established on a statistical basis using the observed energies, directions and times of these events. Assuming many extra-galactic neutrino sources contribute to the total flux, one expects it to appear isotropic and essentially diffuse \cite{Halzen2010}. It is also anticipated to be more likely to produce high energy neutrinos than all atmospheric background components. For example diffusive shock acceleration of protons at astrophysical sources and neutrino production in $pp-$collisions naively predicts a hard energy spectrum $\Phi=dN/dE\sim E^{-2}$ \cite{Halzen2010,Drury1983a}. Standard neutrino oscillations over astronomical baselines equalize any injected neutrino flavor composition at the source towards equilibrium. For example, neutrinos from the pion decay chain have an injected flavor ratio $\left(\Phi_e:\Phi_\mu:\Phi_\tau\right)=(1:2:0)$, which will be observed at Earth as $(1:1:1)$ \cite{Huemmer2010}. Thus, in most IceCube analyses, the flux of cosmic neutrinos is parametrized as single power law with equal contribution from all neutrino flavors 
\begin{align}
\Phi_{\nu_i+\bar{\nu}_i}\left(E\right) / (10^{-18}\,\textnormal{GeV}^{-1}\,\textnormal{cm}^{-2}\,\textnormal{sr}^{-1}\,\textnormal{s}^{-1})=\phi_0\times\left(\frac{E}{100\,\textnormal{TeV}}\right)^{-\gamma},\qquad i\in\{e,\,\mu,\,\tau\}
\label{eq:single_plaw}
\end{align} 
where the spectral index $\gamma$ and flux normalization per neutrino flavor at $E=100\,\textnormal{TeV}$, $\phi_0$, are free parameters.
A diffuse flux of cosmic neutrinos, consistent with these properties, has been observed by IceCube using two complementary searches \cite{IceCubeCollaboration:2013b,Aartsen2015}. First, by isolating a sample of muon tracks coming from the Northern Sky (including horizontal trajectories) it is possible to use the Earth as shield against the atmospheric muon background to achieve a neutrino purity of $\sim 99.7\%$ dominated by CC $\nu_\mu$ (and $\bar{\nu}_\mu$) interactions \cite{IceCubeCollaboration:2016}. Second, atmospheric muon and neutrino backgrounds from the Southern Sky (c.f. atmospheric neutrino self veto \cite{Schoenert2009,vanSanten2014a,Argueelles2018}) can be supressed by focusing on "starting" events with interaction vertices inside the detector. In this channel astrophysical neutrinos contribute most significantly to cascade events \cite{IceCubeCollaboration:2013b}. Because of the detector containment requirements, the neutrino effective area is smaller compared to searches for muon tracks, but the flux of cosmic neutrinos can be studied throughout the entire sky and at energies below the sensitiviy of track searches.
In addition to energy and declination information, astrophysical neutrino candidates can be identified through pointing and time requirements. For example, a clustering of neutrino events above background expectations and consistent with a neutrino point source in the sky or neutrino detections in spatial and temporal coincidence with transient astrophysical phenomena are strong indicators of a cosmic origin \cite{IceCubeCollaboration2019,IceCubeCollaboration2018a}. Finally, the observation and identification of $\nu_\tau$ events would very strongly imply that these are astrophysical in nature, since atmospheric $\nu_\tau$ production is subdominant. Identification of $\nu_\tau$ CC events is challenging and relies on the dominant energy loss pattern: a first cascade at the interaction vertex and a second one from the subsequent decay of the $\tau^{\pm}$ lepton after propagation. If the two cascades are not well separated, the event can not be distinguished from a single cascade \cite{Aartsen2016b}.

\section{Assessing Deep Inelastic Neutrino Interactions}
IceCube is sensitive to the fundamental physics of high energy ($>\,\mathrm{TeV}$) deep inelastic neutrino nucleon scattering (DIS). The interaction cross section can be measured through the absorption of neutrinos in the Earth \cite{Aartsen2017e} and kinematics can be assessed through the reconstruction of the inelasticity from starting $\nu_\mu$-CC events \cite{Aartsen2019}. Fig. \ref{fig:absorption} shows the probability that a neutrino will pass through the Earth as a function of neutrino energy and zenith angle assuming the Standard Model cross section calculation from Cooper-Sarkar et al. \cite{Cooper-Sarkar2011}. For vertical directions significant absorption sets in above $\sim 40\,\mathrm{TeV}$. Potential changes in the $\nu N$-cross section will alter these transmission probabilities and hence enable measurement of the cross section. The inelasticity $y=E_{\textnormal{had}} / E_{\nu} = E_{\textnormal{had}} / (E_\mu + E_{\textnormal{had}})$ in $\nu_\mu$-CC interactions is measured by separately reconstructing the energy of the hadronic cascade $E_{\textnormal{had}}$ and the outgoing muon $E_\mu$ for starting track events. Precise determination of the inelasticity can be used to constrain charm production in $\nu N$ DIS and may enable future determination of the $\nu:\bar{\nu}$ ratio in the atmospheric neutrino flux at energies $<10\,\mathrm{TeV}$ \cite{Aartsen2019}.

\section{Results}
\subsection{Measurements of the Diffuse Astrophysical Neutrino Flux}
\begin{figure}
  \centering
    \includegraphics[width=0.42\textwidth]{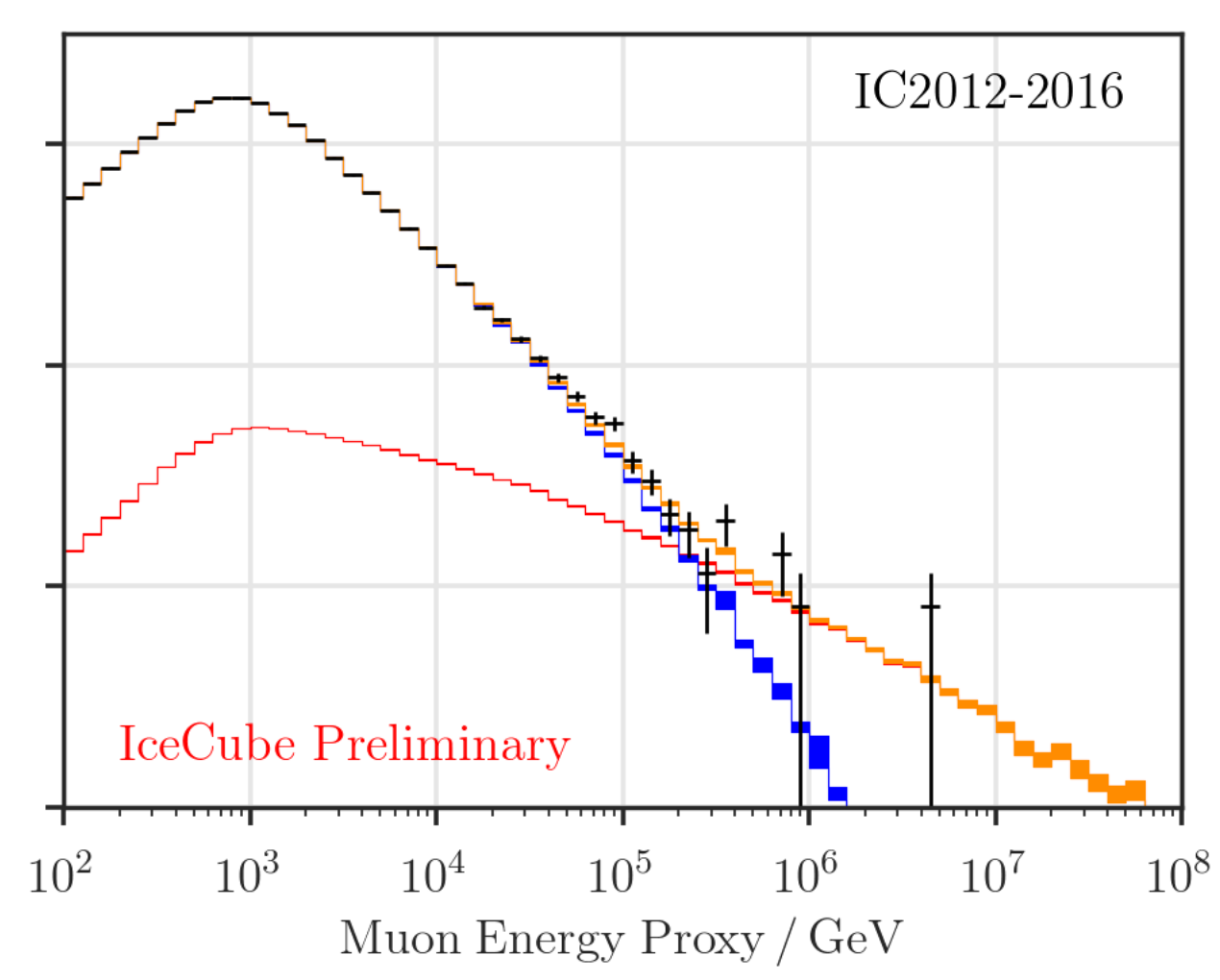}
    \includegraphics[width=0.48\textwidth]{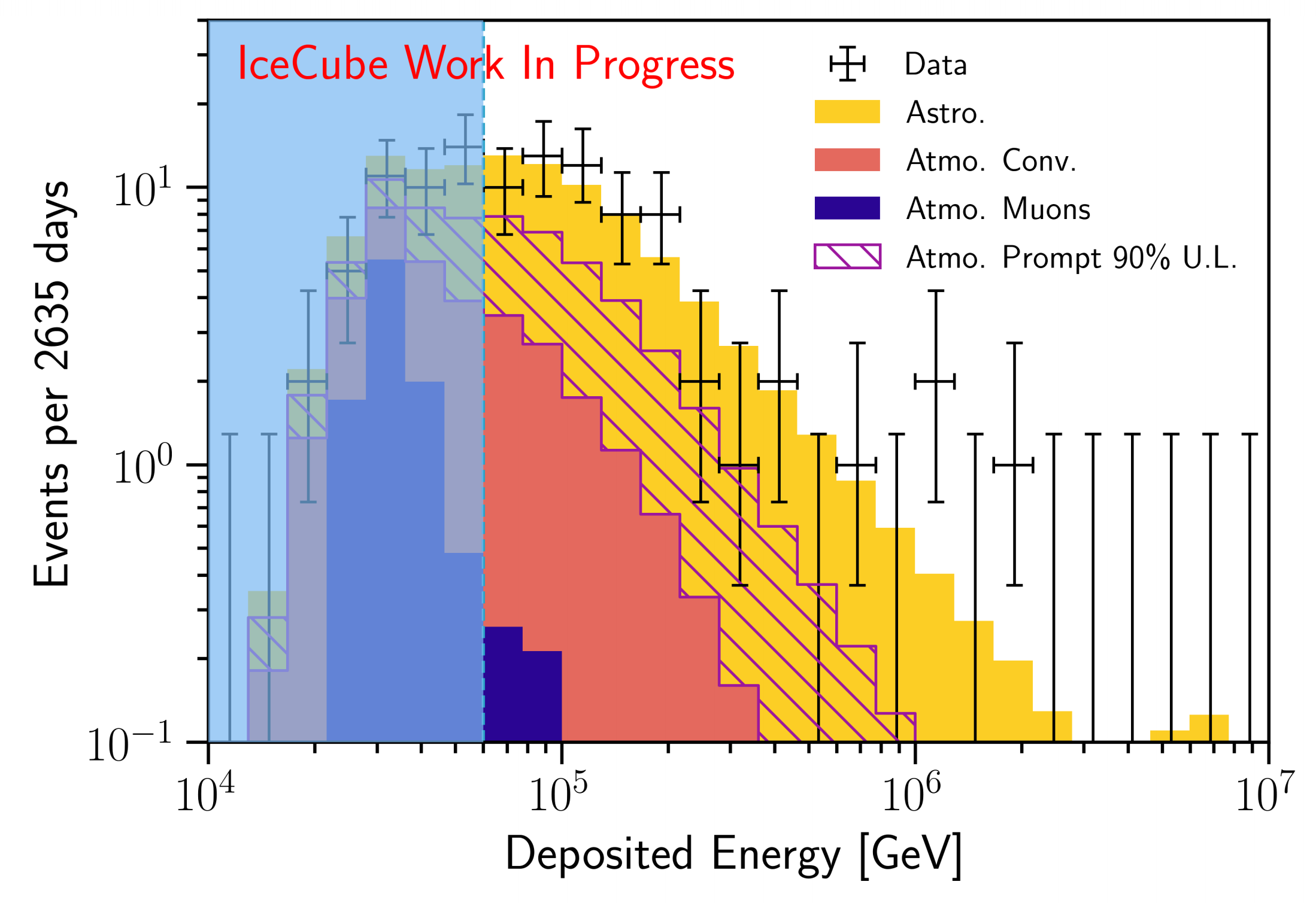}
      \caption{\textbf{Observed energy spectra:} number of observed through-going tracks (left) and high energy starting events (right) as function of estimated event energy. Expectations for atmospheric and astrophysical neutrino components are plotted according to the respective best-fit. Astrophysical neutrinos dominate the high energy regime in both samples.}
      \label{fig:spectra}
\end{figure}
Currently, the sample of through-going tracks from the Northern Sky consists of about $500,000$ $\nu_\mu$-CC events collected during eight years of IceCube operations ($2009-2017$) \cite{Haack2017}. Fig. \ref{fig:spectra} (left) shows the observed energy spectrum using data from the most recent detector configuration ($2012-2016$). At the highest energies the data are clearly inconsistent with atmospheric expectations (blue) and the combined dataset favors an astrophysical neutrino component (red) at a significance of $\sim 6.7\sigma$. The brightest event deposited an energy of $2.6\pm 0.3\,\mathrm{PeV}$ in the detector, which corresponds to a (median) neutrino energy of $\sim 9\,\mathrm{PeV}$.  Assuming a single power law, eq. \eqref{eq:single_plaw}, the cosmic flux is measured above $E_\nu=119\,\mathrm{TeV}$ with best-fit spectral index $\gamma=2.19\pm0.10$ and normalization $\phi_0=1.01^{+0.26}_{-0.23}$ \cite{Haack2017}. The sample of high energy starting events consists of $102$ events ($7.5$ years of data taking), of which $60$ events have reconstructed energies above $60\,\mathrm{TeV}$, see Fig. \ref{fig:spectra} (right) \cite{IceCubeCollaboration:2017b}. This subset of events is used in the statistical analysis, because of the negligible contribution from atmospheric muons. The highest energy event is a $2.0^{+0.3}_{-0.2}\,\mathrm{PeV}$ particle shower. The measurement finds a best-fit spectral index $\gamma=2.91^{+0.33}_{-0.22}$ with normalization $\phi_0=2.19^{+1.10}_{-0.55}$ \cite{IceCubeCollaboration:2017b}. This result favors a flux that is softer than the track-based measurement, but both analyses are consistent in the overlapping energy range. Measurements at even lower energies, down to a few $\mathrm{TeV}$, using starting events and cascades prefer a spectral index of $\gamma\sim 2.5$\cite{IceCubeCollaboration:2015c,Hans}. More statistics are needed to clarify whether hints of possible structure in the flux are real and, for example, due to contributions from several different (astro)physical processes. For this purpose, a global analysis of all IceCube detection channels is currently under development \cite{Nancy2017}. It will include a new selection of partially contained cascades starting at or beyond IceCube's detector boundary \cite{LuLu}. As part of this work, a partially contained cascade event ($E\sim6\,\mathrm{PeV}$), potentially produced through the Glashow resonance interaction, has been observed and is currently under investigation. Finally we searched the sample of high energy starting events for signatures of $\nu_\tau$ CC events using a newly developed double cascade identifier \cite{Stachurska2019}. Two events pass the double cascade selection criteria. Both were scrutinized and one was found to be consistent with two cascade-like energy deposits ($E_1\sim9\,\mathrm{TeV}$, $E_2\sim80\,\mathrm{TeV}$) spatially separated by $17\,\mathrm{m}$. Taking into account this new identifier, the IceCube data remain consistent with equal contributions from all neutrino flavors, but a non-zero $\nu_\tau$ flux cannot firmly be established \cite{Stachurska2019}.

\subsection{Searches for Point Sources of Astrophysical Neutrinos}
\begin{figure}
  \centering
    \includegraphics[width=0.9\textwidth]{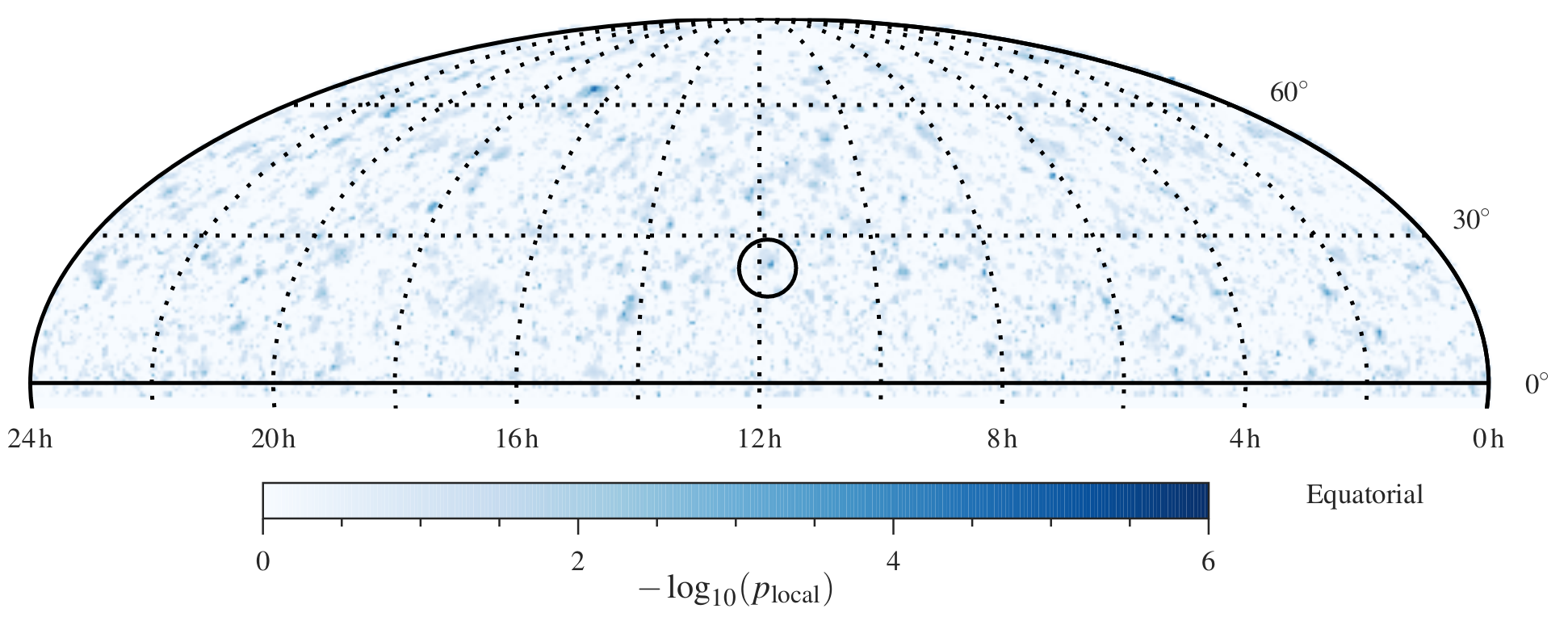}
   \caption{\textbf{IceCube map of Northern Sky:} (local) significance of search for time-integrated, point-like neutrino emission using through-going track events as function of possible source position. see text for details.}
   \label{fig:skymap}
\end{figure}

The sample of through-going tracks from the Northern Hemisphere has been searched for signs of steady, point-like neutrino emission \cite{IceCubeCollaboration2019}. The search relies on an unbinned maximum likelihood method \cite{Braun2008}. Fig. \ref{fig:skymap} visualizes the local p-value obtained as function of the source's position in the sky using a likelihood ratio test that compares a hypothetical point source plus atmospheric and diffuse astrophysical backgrounds against the no-source case. No significant clustering has been found. The most significant position (black circle) corresponds to a trial-corrected p-value of $26.5\%$. The trials factor can be reduced by limiting the search to a smaller set of interesting sky locations selected \textsl{a priori}. The positions of known $\gamma$-ray sources have been selected and studied individually, but no significant signal was found \cite{IceCubeCollaboration2019}. The sensitivity to a set of point sources can be increased by considering the combined signal in what is called a stacking approach. Various subsets of blazars that contribute to the 2nd Fermi-LAT AGN catalogue (2LAC) have been studied but no significant correlations between IceCube neutrinos and these sets of $\gamma$-ray emitting blazars have been observed \cite{Aartsen2017f}. Focusing on HBL\footnote{HBL blazars are those that have the synchrotron peak located at X-ray energies.} blazars included in the 2FHL catalogue, and assuming their energy spectra to follow a single power law with $\gamma=2.13$, the contribution of the HBL blazars to the total diffuse cosmic neutrino flux must be smaller than $\sim 6\%$ ($90\%\,\mathrm{C.L.}$) \cite{Huber2018}. Finally, the absence of spatial and temporal correlation between IceCube neutrinos and GRBs challenges fireball models of prompt GRB emission \cite{Aartsen2017g}.

\subsection{Evidence for Neutrino Emission from Blazar TXS 0506+056}
\begin{figure}
  \centering
    \includegraphics[width=0.9\textwidth]{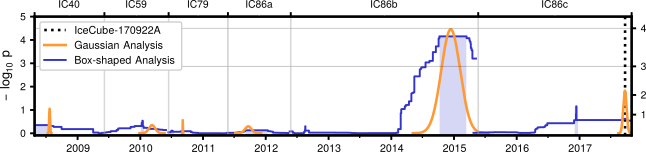}
      \caption{Search for time-dependent neutrino emission from blazar TXS 0506+056 in IceCube data from Apr 5, 2008 to Oct 31, 2017. P-value in each data taking period as function of center of the gaussian (orange) or box-shaped (blue) time window. see text for details.}
      \label{fig:timedep}
\end{figure}

IceCube regularly alerts other experiments in near real-time about interesting neutrino observations in order to enable electromagnetic follow-up observations \cite{Aartsen2017h}. On September 22, 2017, IceCube observed a high-energy  through-going track event (IC-170922A) with sizeable probability of being astrophysical in origin ($56.5\%$) \footnote{Astrophysical origin refers to a $E^{-2}$ diffuse astrophysical neutrino flux. See \cite{Aartsen2017h} for a precise definition of this ''signalness'' measure.} \cite{IceCubeCollaboration2018a}. The most likely energy of this neutrino is $\sim290\,\mathrm{TeV}$. Several experiments studied the region of the sky consistent with the direction of the event. The Fermi-LAT reported the observation of a $\gamma$-ray flare from the blazar TXS 0506+056 (redshift $z=0.3365\pm 0.0010$ \cite{Paiano2018}) during the time of the event \cite{IceCubeCollaboration2018a}. The MAGIC experiment also detected $\gamma$-rays from this source with energies up to $400\,\mathrm{GeV}$ over a 10-day period, starting 2 days after the event  \cite{IceCubeCollaboration2018a}. The chance-coincidence probability of this observation has been studied under several assumptions about the positive correlation between $\gamma$-ray and neutrino fluxes. The observation was found to be rare, suggesting a correlation at the $\sim 3\sigma$ level. Subsequently, the direction of TXS 0506+056 has been searched for time dependent, point-like neutrino emission with IceCube data recorded since 2009 using the method described in Braun et al. \cite{Braun2010}. Time-dependent neutrino emission has been accounted for assuming Gaussian or box-shaped time profiles \cite{science2018}. In both analyses an excess of high energy neutrino events was found in the period from September 2014 to March 2015. In total $13\pm5$ neutrino events are estimated to contribute to this neutrino flare candidate and found to be consistent with a hard spectrum $\gamma=2.1\pm0.2$. Fig. \ref{fig:timedep} shows the p-value of the excess compared to the background-only assumption as function of the (central) time of the flare. The significance of the excess is $3.5\sigma$ in favor of neutrino emission from the direction of TXS 0506+056 over the background-only assumption \cite{science2018}.

\subsection{Measurements of Crosssection and Inelasticity in DIS Interactions}
Using a one year dataset of through-going tracks from the Northern Sky, IceCube has measured the $\nu N$ DIS interaction cross section from $6.3\,\mathrm{TeV}$ to $980\,\mathrm{TeV}$ in units of the Standard Model (SM) cross section calculated by Cooper-Sarkar et al.\cite{Cooper-Sarkar2011}. The measured value $1.30^{+0.21}_{-0.19}\,(\mathrm{stat.})\,^{+0.39}_{-0.43}\,(\mathrm{syst.})$ is consistent with the SM expectation of $1.00$ in these units \cite{Aartsen2017e}. Similar conclusions are drawn from complementary cross section analyses using high energy starting events \cite{tianlu} and high energy cascades \cite{Xu2018a}. These measurements concern energies more than one order of magnitude beyond the reach of previous accalerator-based measurements. Using a new event selection optimized for the identification and characterization of starting track events, IceCube also studied the inelasticity $y$ in $\nu_\mu$ CC interactions \cite{Aartsen2019}. The analysis is based on five years of data and found good agreement between data and SM predictions. Through the effect of charm production on the inelasticity distribution at neutrino energies between $1.5\,\mathrm{TeV}$ to $340\,\mathrm{TeV}$, zero charm production in $\nu_\mu$ CC interactions was excluded at $\sim90\%$ C.L. \cite{Aartsen2019}.

\section{Conclusion}
The IceCube experiment has achieved important milestones on the road to routine neutrino astronomy. A diffuse flux of cosmic neutrinos has been observed and a first neutrino (and cosmic ray) source potentially identified. IceCube measurements also contribute to fundamental neutrino physics at energies well beyond the direct reach of accelerator based experiments. In addition the IceCube Collaboration maintains a rich dark matter and neutrino oscillation program not covered in this contribution. An upgrade of the detector has been approved and is intended as a stepping stone towards a next generation particle and astrophysics facility at the South Pole.



\section*{References}

\begin{thebibliography}{99}

\bibitem{Fenu2017}
	PA Collaboration: F. Fenu, et al. \textit{Proceedings of Science}, PoS(ICRC2017)486 (2018)
\bibitem{ivanov2019}
	TA Collaboration: D. Ivanov, et al. \textit{EPJ Web of Conferences} 210, 01001 (2019)
\bibitem{Halzen2002}
	F. Halzen and D. Hooper. \textit{Reports on Progress in Physics} 65, 7, 1025 (2002)
\bibitem{Greisen1966}
	K. Greisen. \textit{Physical Review Letters} 16, 17, 748 (1966)
\bibitem{Zatsepin1966}
	G. T. Zatsepin and V. A. Kuz’min. \textit{Soviet J. of Exp and Theor. Phys. Lett.} 4, 78 (1966).
\bibitem{Gould1967}
	R. J. Gould and G. P. Schroeder. \textit{Physical Review Journal} 155,  5, 1404 (1967).
\bibitem{Nikishov1961}
	A. Nikishov. \textit{Zhur. Eksptl’. i Teoret. Fiz.} 41 (1961)
\bibitem{Dermer2014}
	C. D. Dermer et al. \textit{Journal of High Energy Astrophysics} 3-4, 29 (2014)
\bibitem{Murase2017}
	K. Murase. \textit{Active Galactic Nuclei as High-Energy Neutrino Sources} in \textit{Neutrino Astronomy: Current Status, Future Prospects}, 15, World Scientific (2017)
\bibitem{Padovani2015}
	P. Padovani et al. \textit{Mon. Notices Royal Astron. Soc}, 452, 2, 1877 (2015)
\bibitem{Waxman1997a}
	E. Waxman and J. Bahcall. \textit{Phys. Rev. Lett.} 78, 2292 (1997)
\bibitem{Tamborra2015}
	I. Tamborra and S. Ando. \textit{Journal of Cosmology and Astroparticle Physics}, 9, 36 (2015)
\bibitem{Ackermann2019}
	M. Ackermann et al. \textit{arXiv e-prints}, arXiv:1903.04333 (2019)
\bibitem{Halzen2010}
	F. Halzen and S. Klein. \textit{Review of Scientific Instruments}, 81, 081101 (2010)
\bibitem{Aartsen2017d}
	IceCube Collab.: M. G. Aartsen et al. \textit{Journal of Instrumentation} 12, 3, 03012 (2017)
\bibitem{IceCubeCollaboration2019}
	IceCube Collab.: M. G. Aartsen et al. \textit{The European Physical Journal C}, 79, 234 (2019)
\bibitem{IceCubeCollaboration:2014}
	IceCube Collab.: M. G. Aartsen et al. \textit{Journal of Instrumentation} 9, 3, 03009 (2014)
\bibitem{Gaisser2002}
	T. K. Gaisser and M. Honda. \textit{Ann. Rev. of Nuclear and Particle Science} 52, 1, 153 (2002)
\bibitem{Drury1983a}
	L. O. Drury. \textit{Reports on Progress in Physics}, 46, 973 (1983)
\bibitem{Huemmer2010}
	S. Huemmer et al. \textit{Astroparticle Physics}, 34, 4, 205 (2010)
\bibitem{IceCubeCollaboration:2013b}
	IceCube Collab.: M. G. Aartsen et al. \textit{Science}, 342, 6161, 1242856 (2013)
\bibitem{Aartsen2015}
	IceCube Collab.: M. G. Aartsen et al. \textit{Physical Review Letters}, 115, 8, 081102 (2015)
\bibitem{IceCubeCollaboration:2016}
	IceCube Collab.: M. G. Aartsen et al. \textit{The Astrophysical Journal}, 833, 1, 3 (2016)
\bibitem{Schoenert2009}
	S. Schoenert et al. \textit{Physical Review D}, 79, 4, 04300 (2009)
\bibitem{vanSanten2014a}
	J. van Santen et al. \textit{Physical Review D}, 90, 2, 023009 (2014)
\bibitem{Argueelles2018}
	C. A. Arguelles, \textit{J. Cosmol. Aastropart. Phys.}, 7, 047 (2018)
\bibitem{IceCubeCollaboration2018a}
	IceCube Collab.: M. G. Aartsen et al. \textit{Science}, 361, 6398, 1378 (2018)
\bibitem{Aartsen2016b}
	IceCube Collab.: M. G. Aartsen et al. \textit{Physical Review D}, 93, 2, 022001 (2016)
\bibitem{Aartsen2017e}
	IceCube Collab.: M. G. Aartsen et al. \textit{Nature}, 551, 7682, 596 (2017)
\bibitem{Aartsen2019}
	IceCube Collab.: M. G. Aartsen et al. \textit{Physical Review D}, 99, 3, 032004 (2019)
\bibitem{Cooper-Sarkar2011}
	A. Cooper-Sarkar et al. \textit{Journal of High Energy Physics}, 08, 042 (2011)
\bibitem{Haack2017}
	IceCube Collab.: C. Haack, et al. \textit{Proceedings of Science}, PoS(ICRC2017)1005 (2018)
\bibitem{IceCubeCollaboration:2017b}
	IceCube Collab.: C. Kopper, et al. \textit{Proceedings of Science}, PoS(ICRC2017)981 (2018)
\bibitem{IceCubeCollaboration:2015c}
	IceCube Collab.: M. G. Aartsen, et al. \textit{Physical Review D}, 91, 2, 022001 (2015)
\bibitem{Hans}
	IceCube Collab.: H. Niederhausen, et al. \textit{Proc. of Science}, PoS(ICRC2017)968 (2018)
\bibitem{Nancy2017}
	IceCube Collab.: N. Wandkowsky, et al. \textit{Proceedings of Science}, PoS(ICRC2017)976 (2018)
\bibitem{LuLu}
	IceCube Collab.: L. Lu, et al. \textit{Proceedings of Science}, PoS(ICRC2017)1002 (2018)
\bibitem{Stachurska2019}
	IceCube Collab.: J. Stachurska, et al. \textit{Proceedings of Science}, PoS(ICRC2019)1015 (2019)
\bibitem{Braun2008}
	J. Braun et al. \textit{Astroparticle Physics}, 29, 4, 299 (2008)
\bibitem{Aartsen2017f}
	IceCube Collab.: M. G. Aartsen et al. \textit{Astrophysical Journal}, 835, 1, 45 (2017)
\bibitem{Huber2018}
	IceCube Collab.: M. Huber, et al. \textit{Proceedings of Science}, PoS(ICRC2017)994 (2018)
\bibitem{Aartsen2017g}
	IceCube Collab.: M. G. Aartsen et al. \textit{Astrophysical Journal}, 843, 2, 112 (2017)
\bibitem{Aartsen2017h}
	IceCube Collab.: M. G. Aartsen et al. \textit{Astroparticle Physics}, 92, 30 (2017)
\bibitem{Paiano2018}
	S. Paiano et al. \textit{Astrophysical Journal Letters}, 854, 2, 32 (2018)
\bibitem{Braun2010}
	J. Braun et al. \textit{Astroparticle Physics}, 33, 3, 175  (2010)
\bibitem{science2018}
	IceCube Collab.: M. G. Aartsen et al. \textit{Science}, 361, 6398, 147 (2018)
\bibitem{tianlu}
	IceCube Collab.: T. Yuan, et al. \textit{Proceedings of Science}, PoS(ICRC2019)1040 (2019)
\bibitem{Xu2018a}
	IceCube Collab.: Y. Xu, et al. \textit{XXVI International Workshop on Deep-Inelastic Scattering and Related Subjects}, Kobe. (2018)
\end{thebibliography}
\end{document}